\documentclass{article}
\usepackage{amsmath}
\usepackage[dvips]{graphicx}
\begin{document}
\title{ 
Supersymmetric Dark Matter with GLAST
}
\author{
Andrea Lionetto        \\
{\em Dipartimento di Fisica e INFN Roma Tor Vergata} \\
{\em Via della Ricerca Scientifica, 1 Roma} \\
}
\maketitle
\baselineskip=11.6pt
\begin{abstract}
We present a short review of the status of the dark matter problem. In particular we show that one of the best motivated candidate for the dark matter is the neutralino, a supersymmetric particle. Finally we study the possibility to detect $\gamma$-rays coming from neutralino pair annihilations that take place in our Galactic Center with the upcoming satellite detector GLAST.
\end{abstract}
\baselineskip=14pt
\section{Introduction}
The dark matter problem is one of the most fascinating and intriguing issue 
in the astroparticle physics. This subject requires theoretical concepts and 
experimental techniques coming from cosmology, astrophysics and fundamental 
particle physics.

By definition dark matter is a kind of non luminous matter. Its existence is required
 by a bunch of striking experimental evidences as well as by some compelling theoretical
motivations. One of most interesting experimental evidences is given by the observations of
the spiral galaxy rotation curves~\cite{Griest:kj}. They can be determined well outside the luminous core 
of the galaxy, measuring the circular velocities of the neutral hydrogen clouds using 
the $\lambda=21$ cm lines emissions. The result when expressed as a function 
of the galactocentric distance $R$ is that $v_C\sim \textrm{const.}$ for large $R$ (in particular for
$R>R_{lum}\simeq 5\,\textrm{Kpc}$) rather than the expected $v_C\propto R^{-1/2}$. If we suppose 
that the first Newton's law still holds at the galactic scale we must conclude that $M(<R)\propto R$
where $M(<R)$ is the total mass inside a shell of radius $R$. Hence there must be a ``dark'' contribution to the galactic mass $M$.

Recent measurements~\cite{wmap} of the anisotropy of the Cosmic Microwave Background (CMB) strongly indicates that the total matter contribution is $\Omega h^2\sim 0.3$ where $h\sim 0.71$ is the Hubble constant in units of 100 km s$^{-1}$ Mpc$^{-1}$. Hence $\Omega_{m} h^2$ is much larger than the baryonic term $\Omega_{b}h^2\sim 0.02$. Necessarily the dominant matter component must be of a non baryonic form. It turns out that the best motivated candidates are some kind of Weak Interacting Massive Particles (WIMPs). Usually these type of candidates are stable particles that appear in various extension of the Standard Model (SM) of particle physics.

\section{Supersymmetric dark matter}
From the theoretical point of view the most interesting extension of the SM involves supersymmetry. Here we will focus in particular on $N=1$ supersymmetric extension of the SM with soft supersymmetry breaking terms. This is the so called Minimal Supersymmetric Standard Model (MSSM)~\cite{Haber:1984rc}. Usually the Lightest Supersymmetric Particle (LSP) that appear in the MSSM mass spectrum is a good CDM candidate~\cite{Griest:kj}. It turns out that the LSP is often the lightest neutralino. The four neutralino gauge eigenstates are linear combination of the gaugino and higgsino fields:
\begin{equation}
\chi_i=N_{i1}\tilde{B}+N_{i2}\tilde{W}^0+N_{i3}\tilde{H}_d^0+N_{i4}\tilde{H}_u^0
\end{equation}
where $i=1,\ldots 4$. The corresponding mass eigenstates can be obtained by diagonalizing the mass matrix:
\begin{equation}
\bf{M_{\tilde{N}}}=\left (\begin{array}{c c c c}
M_1 & 0 & -c_\beta s_W m_Z & s_\beta s_W m_Z \\
0 & M_2 & c_\beta c_W m_Z & -s_\beta c_W m_Z \\
-c_\beta s_W m_Z & c_\beta c_W m_Z &0 & -\mu \\
s_\beta s_W m_Z & -s_\beta c_W m_Z & -\mu & 0
\end{array}\right )
\label{neutmassmatrix}
\end{equation}
that depends by various MSSM parameters, like the gaugino masses $M_1$ and $M_2$ (that are associated, in the lagrangian, to explicit supersymmetry breaking terms) or the $\mu$ parameter that appears in the Higgs sector~\cite{Martin:1997ns}. The lowest eigenvalue of the mass matrix in equation~\ref{neutmassmatrix} is exactly the LSP, until the so called R-parity is conserved~\cite{Martin:1997ns}.
The neutralino is a Majorana fermion, so it coincides with its own antiparticle, and it is a weak interacting particle. The R-parity conservation implies that the lightest neutralino cannot decay in SM particles and so it is a stable particle. For these reasons the neutralino is a good CDM candidate. 
Moreover a neutralino pair can self annihilate into various SM particles that we can detect.

\section{Supersymmetric dark matter with GLAST}
In the previous section we have seen that the neutralino is a good CDM candidate. It is possible to study its properties through indirect detection of cosmic rays coming from pair annihilations in the dark galactic halo. In particular here we focus on the cosmic $\gamma$-rays from the Galactic Center (GC).

The EGRET data from the GC \cite{Mayer} give a strong indication of an excess with respect of the ``standard'' model production of $\gamma$-rays \cite{smapj}. In the standard picture, the diffuse $\gamma$-ray background is due to the following processes:
\begin{itemize}
\item $\pi^0$ production that promptly decays in $2\gamma$
\[
p+X\to..\to\pi^{0}\to2\gamma,\qquad He+X\to..\to\pi^{0}\to2\gamma
\]
where $p$ and $He$ are primary protons and heliums and $X$ is the interstellar hydrogen or helium.
\item Bremsstrahlung
\item Inverse Compton
\end{itemize}
We are only interested in the energy range above $E>1$ GeV where the dominant component is by far the $\pi^0$ production. In the subsequent analysis we consider only this component for the diffuse background.

\begin{figure}[t]
\begin{center}
\includegraphics[scale=0.45]{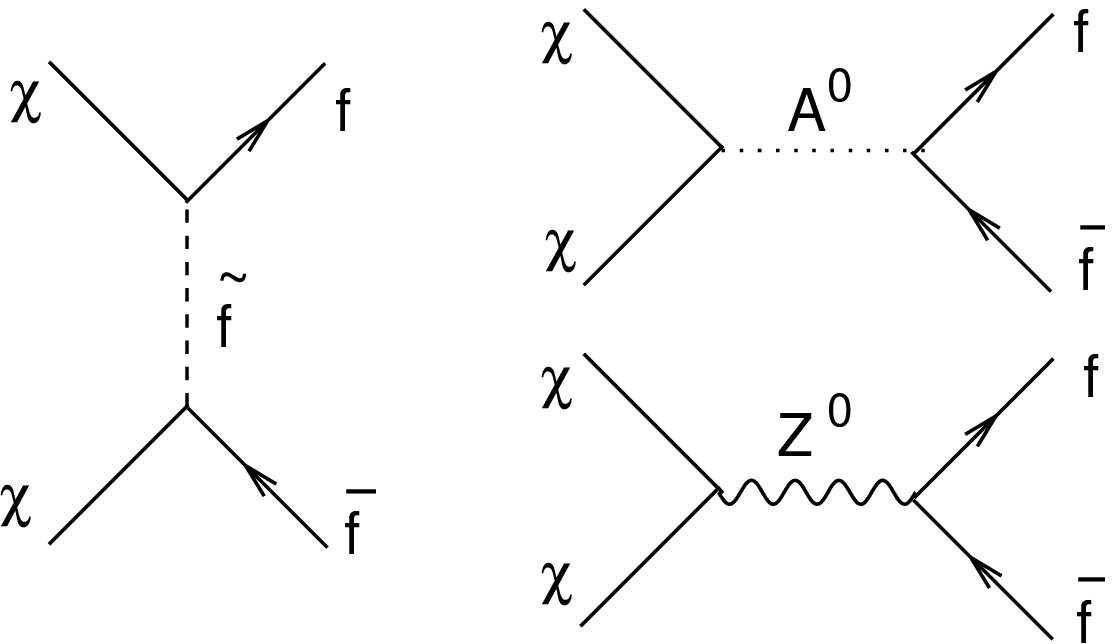}
\includegraphics[scale=0.45]{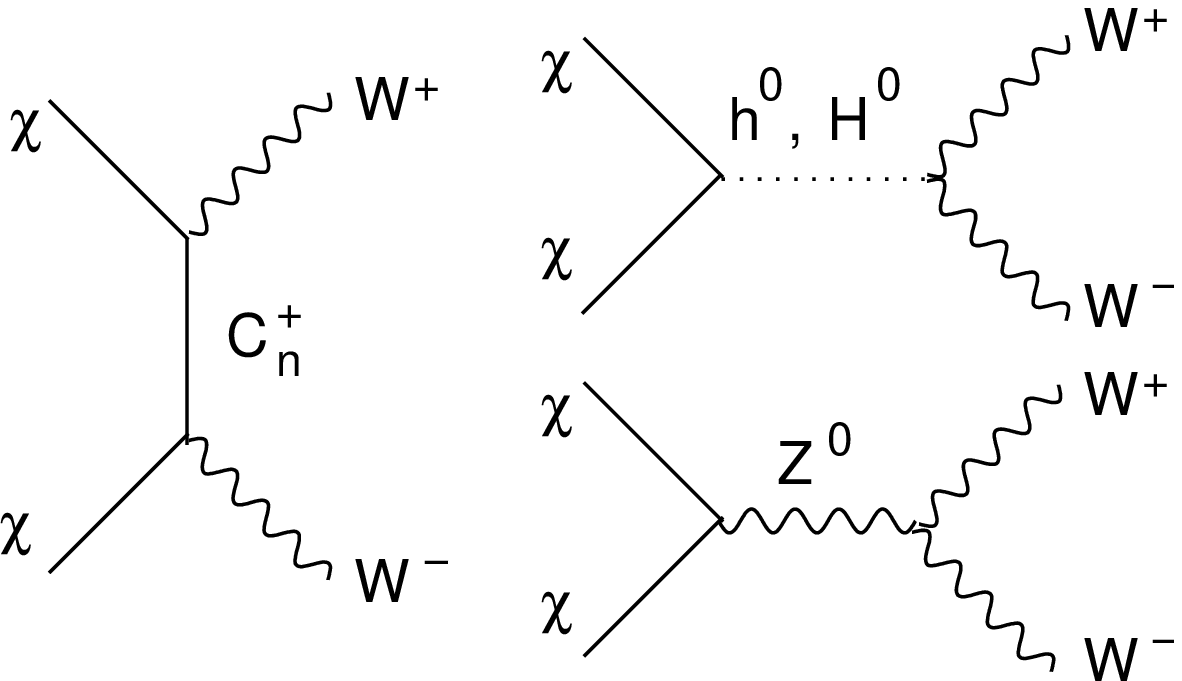}
\includegraphics[scale=0.45]{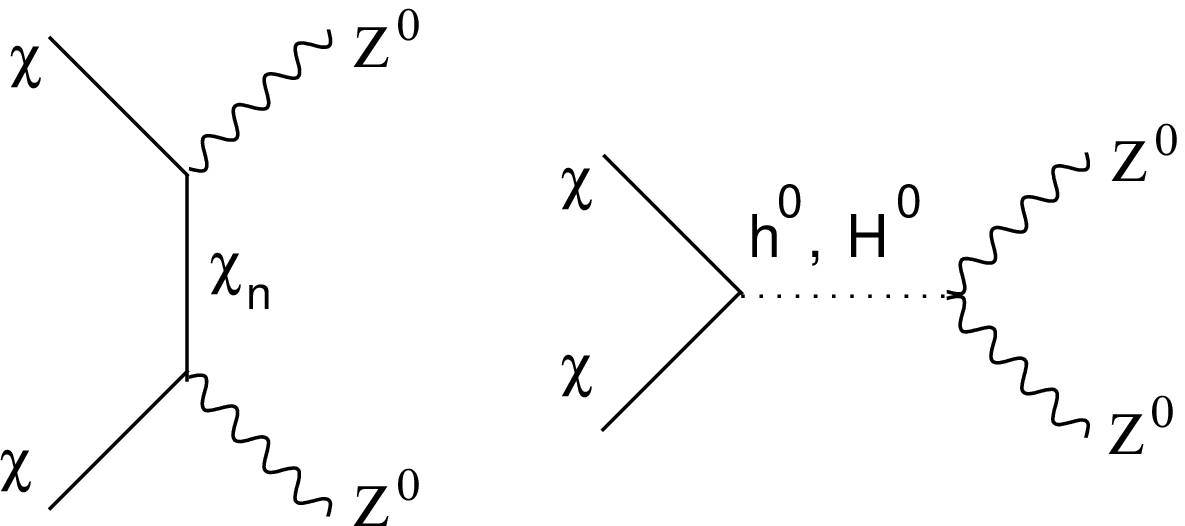}
\caption{Relevant intermediate annihilation channels for a Majorana fermion WIMP (in particular for a neutralino).}
\label{feynman-diag}
\end{center}
\end{figure}

We want to explain the GC excess adding a neutralino induced component in the $\pi^0$ production. The leading intermediate annihilation channels for the $\pi^0$ production, through fragmentation and/or decay processes, are often $b\bar{b}$, $t\bar{t}$, $W^+ W^-$ and $Z^0 Z^0$ (see figure \ref{feynman-diag}). This is valid not only for the neutralino but also for a generic Majorana fermion WIMP, as for such particles the s-wave annihilation rate into lighter fermions is suppressed by a factor $m_f^2 /m_\chi ^2$, where $m_f$ is the final state fermion mass and $m_\chi$ is the WIMP mass.

We can write the total diffuse $\gamma$-ray flux as a sum of two components:
\begin{equation}
\phi_\gamma^{TOT}=\phi_\gamma^{BKG}+\phi_\gamma^{\chi\chi}=N_b S_b+N_\chi \phi_\chi
\end{equation}
where the background flux $\phi_\gamma^{BKG}$ is splitted into two factors:
\begin{eqnarray}
S_{b}(E_\gamma) & = & \frac{1}{\left( 1\,cm^2 sr\right)}\cdot Em(E_\gamma),\nonumber\\
N_b & = & \frac{1}{\left(1\; \rm{cm}^{-2}\rm{sr}^{-1}\right)}\cdot
  \int_{l.o.s.} dl \frac{n_H(l)}{4\pi}
  \frac{\phi^{prim}_{p}(l)}{\phi^{prim}_{p}(l=0)}
\end{eqnarray}
where $Em(E_\gamma)$ (measured in $\textrm{GeV}^{-1}\, \textrm{s}^{-1}$) is the local emissivity per hydrogen atom. The neutralino induced component $N_\chi \phi_\chi$ is given by~\cite{Cesarini:2003nr}:
\begin{eqnarray}
\phi_\chi (E_\gamma) & = & 3.74 \cdot 10^{-10}\left( \frac{\sigma_{ann}\,v}{10^{-26}\;
  \rm{cm}^3 \rm{s}^{-1}}\right)\left( \frac{50\; \rm{GeV}}{m_\chi}\right)^2
  \sum_f \frac{dN_f}{dE} B_f\nonumber\\
N_\chi & = & \left<J(\psi)\right>_{\Delta\Omega}=\frac{1}{\Delta\Omega}\int_{\Delta\Omega}d\Omega' J(\psi')
\end{eqnarray}
where $\Delta\Omega$ is the detector angular acceptance and where we have introduced the dimensionless function J that contains the dependence on the dark halo density profile:
\begin{equation}
J(\psi)=\frac{1}{8.5 Kpc} \left(\frac{1}{0.3 GeV/cm^{3}}\right)\int_{l.o.s} \rho_{\chi}^2 (l)dl(\psi)
\end{equation}
We have introduced the unknown normalization factor $N_b$ in order to take into account our ignorance of the exact hydrogen column density, \emph{i.e.} the interstellar medium, while the $N_\chi$ factor parametrize our ignorance of the dark matter halo model. Results from N-body simulations~\cite{nfw}\cite{moore2} give for for the dark matter density:
\begin{equation}
\rho(r)=\rho_0 \left(\frac{r_0}{r}\right)^\gamma \left[\frac{1+(r_0/a)^\alpha}{1+(r/a)^\alpha}\right]^{\left(\beta -\gamma\right)/\alpha}
\end{equation}
where $\alpha$, $\beta$ and $\gamma$ are parameters that describe the halo profile (Isothermal sphere, NFW, Moore, ..), $\rho_0$ is the WIMP density measured here and $r_0$ is the galactocentric distance. 

\begin{figure}[t]
\begin{center}
\includegraphics[scale=0.3]{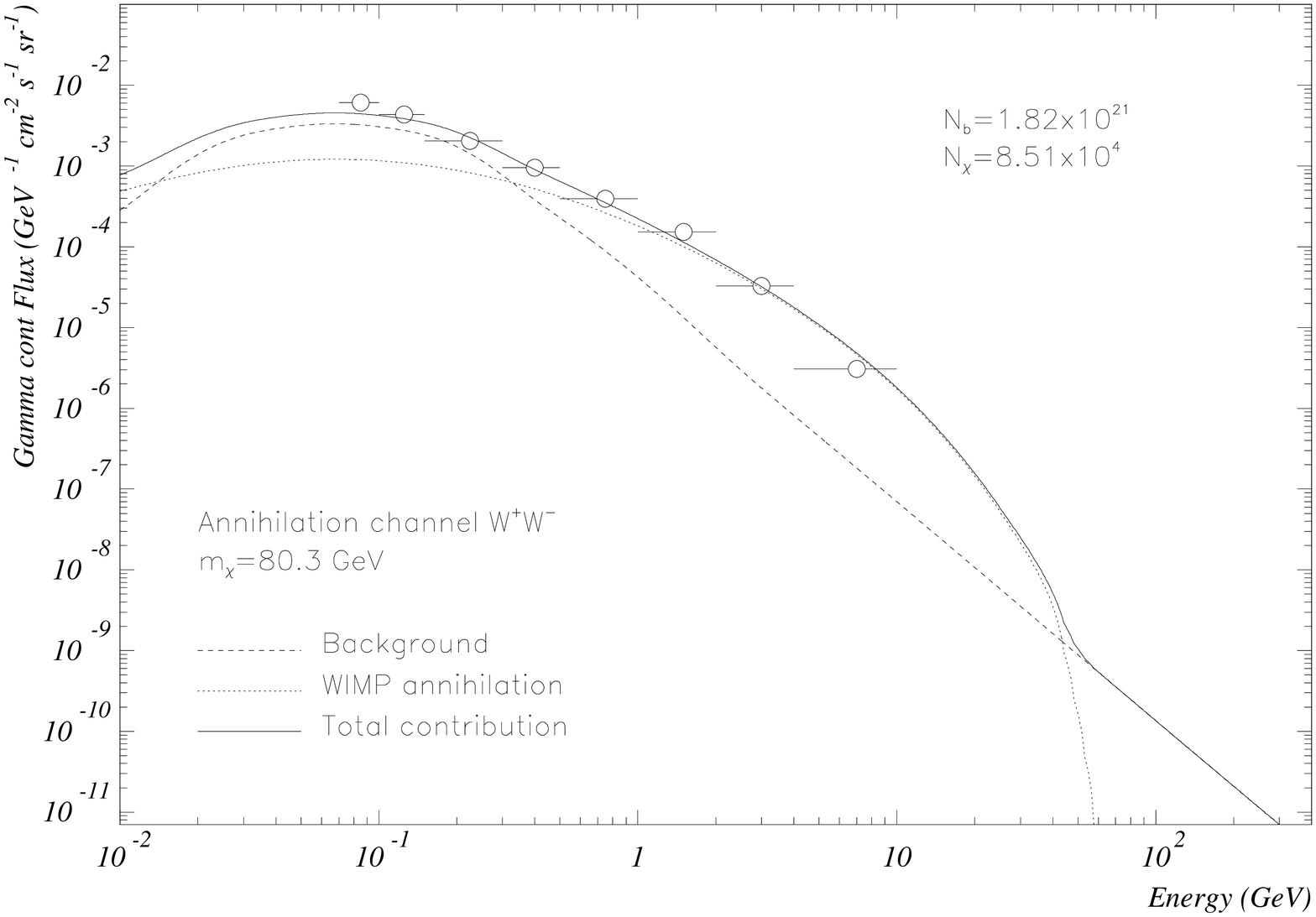}
\caption{Fit of the GC EGRET data assuming the the excess is due to a WIMP (of mass $m_\chi =80$ GeV) induced component with $W^+ W^-$ dominant annihilation channel. The best fit parameters $N_b$ and $N_\chi$ are indicated in the upper right corner.}
\label{egretdata-fit}
\end{center}
\end{figure}

\begin{figure}[t]
\begin{center}
\includegraphics[scale=0.3]{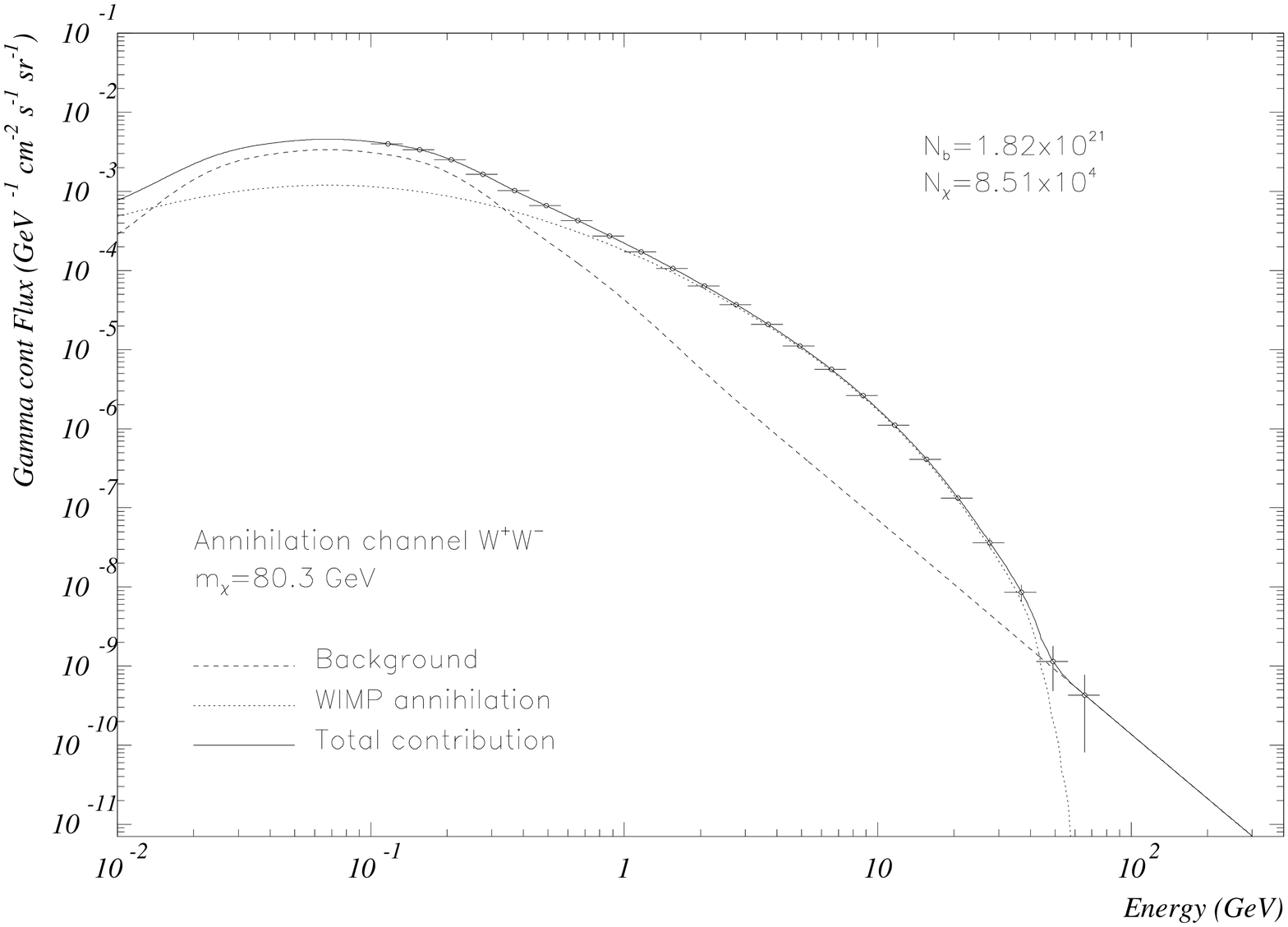}
\caption{Data set collected by GLAST in 2 years in the case the EGRET GC excess is due to a WIMP induced flux.}
\label{egretdata-glast}
\end{center}
\end{figure}

Having introduced the neutralino induced component the fit of the GC EGRET data greatly improves. In the context of a simplified \emph{toy model}~\cite{Cesarini:2003nr}, valid for a generic Majorana fermion WIMP of mass $m_\chi$ and with a fixed dominant intermediate annihilation channel, the result is shown in figure~\ref{egretdata-fit}.
Our analysis shows that the EGRET data fit improves for small neutralino masses. In order to extract much more information about the GC EGRET excess we have to wait for the upcoming generation of telescope satellite, like GLAST~\cite{glast}. With respect to EGRET, GLAST has a wider energy range, increased effective area and better energy and angular resolution. Hence we have studied what kind of data GLAST would collect from the GC, under the hypotesis that the excess, as mapped by EGRET, is due to WIMP annihilations. Relying on a simplified picture of the GLAST detector performances (energy resolution of about $10\%$, angular resolution of $10^{-5}$ sr and peak effective area of $\sim 10^4\, \textrm{cm}^2$), we have computed the GC data set which will be obtained by GLAST in 2 years. We have superimposed the error bars associated to the statistical errors only for the chosen energy binning and with an angular acceptance of $\Delta\Omega=10^{-3}\,\textrm{sr}$. The result, in this case, is shown in figure~\ref{egretdata-glast}.

All the previous results are valid in the case of a generic WIMP. At the end we want to identify the neutralino as our WIMP candidate. To do so we must consider a ``realistic'' supersymmetric model, \emph{i.e.} the MSSM. In order to reduce the number of free parameters of the MSSM it is possible different supersymmetry breaking scenarios in which the supersymmetry breaking terms derive from an underlying high energy theory. One of the most widely studied frameworks is the so-called minimal supergravity (mSUGRA) or constrained MSSM (cMSSM)~\cite{msugra}. 
In the mSUGRA model there is an important assumption about the \emph{universality} of all the coupling constants at the grand unification scale (GUT). With these constraints the number of free parameters is only five:
\[
m_{1/2},\, m_0,\, sign(\mu),\, A_0,\, tan\beta
\]
where $m_0$ is the common scalar mass, $m_{1/2}$ is the common gaugino mass and $A_0$ is the proportionality factor between the supersymmetry breaking trilinear couplings and the Yukawa couplings. $\tan\beta$ denotes the ratio of the vacuum
expectation values of the two neutral components of the SU(2) Higgs doublet, while
the Higgs mixing $\mu$ is determined (up to a sign) by imposing the Electro-Weak Symmetry 
Breaking (EWSB) conditions at the weak scale. In this context 
the MSSM can be regarded as an effective low energy theory. The parameters at the weak energy scale 
are determined by the evolution of those at the unification scale, according to the renormalization 
group equations (RGEs).
Fixing $\tan\beta$, $A_0$ and $sign(\mu)$ we have performed an accurate scan in the $(m_0,m_{1/2})$ mSUGRA parameter space, looking for the minimum normalization factor $N_\chi$ needed to be able to single out the neutralino annihilation signal with GLAST.
In this case we have taken advantage of the better GLAST angular resolution ($\Delta\Omega=10^{-5}\,\textrm{sr}$) with respect to EGRET. 
In this deeper analysis we have also computed the neutralino relic density:
\begin{equation}
\Omega_\chi h^2=\frac{m_\chi n_\chi}{\rho_c}
\end{equation}
solving the Boltzmann equation that describes the time evolution of the neutralino\footnote{valid also for a generic WIMP} number density $n_{\chi}(t)$:
\begin{equation}
\frac{dn_{\chi}}{dt}+3H n_{\chi}=
-\left<\sigma_{ann}\,v\right>\left[\left(n_{\chi}\right)^2-\left(n_{\chi}^{eq}\right)^2\right]
\end{equation}
This equation can be easily solved numerically including resonances, threshold effects and all possible coannihilation processes~\cite{newcoann}.

\begin{figure}[t]
\begin{center}
\includegraphics[scale=0.4]{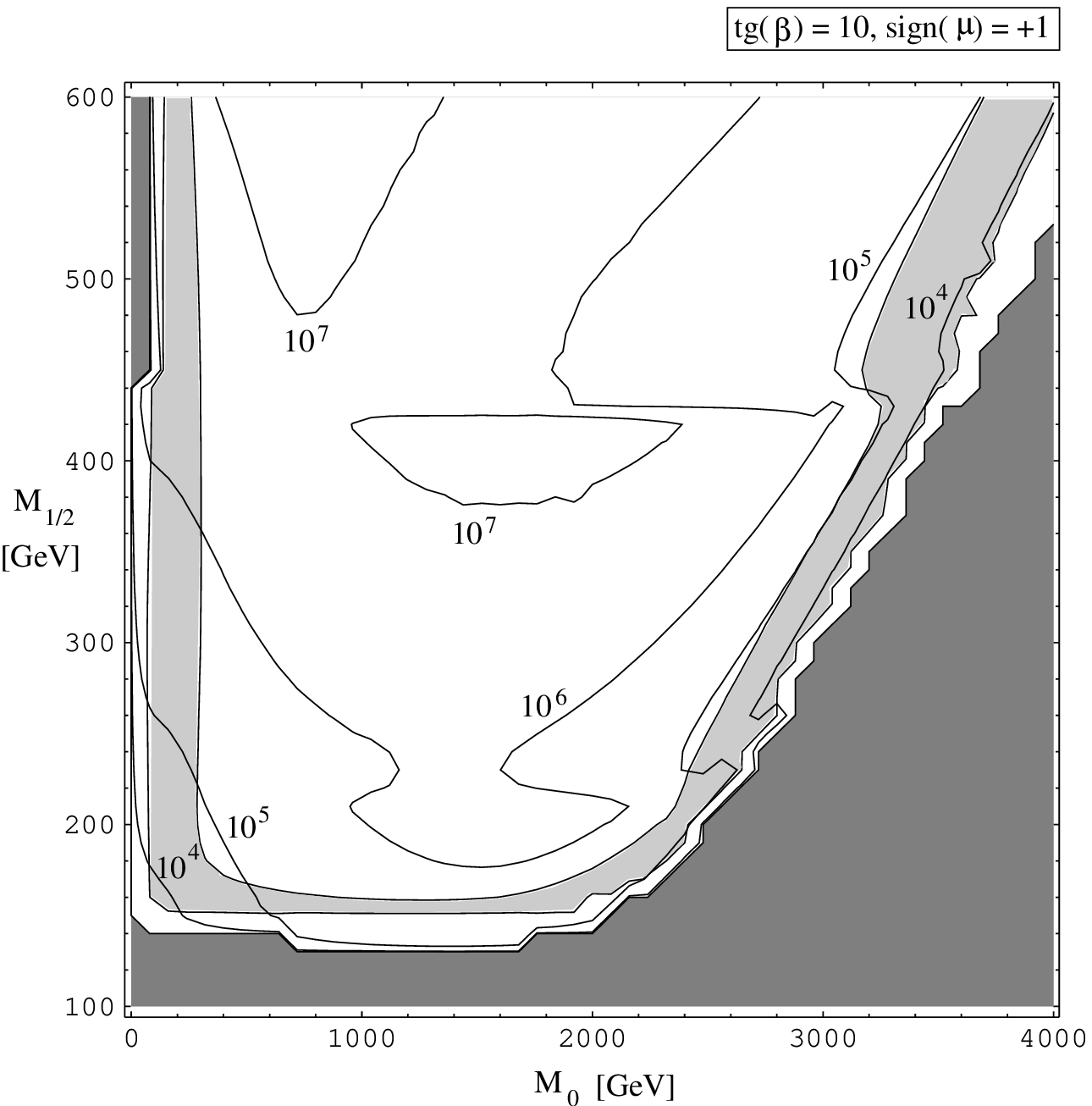}
\includegraphics[scale=0.4]{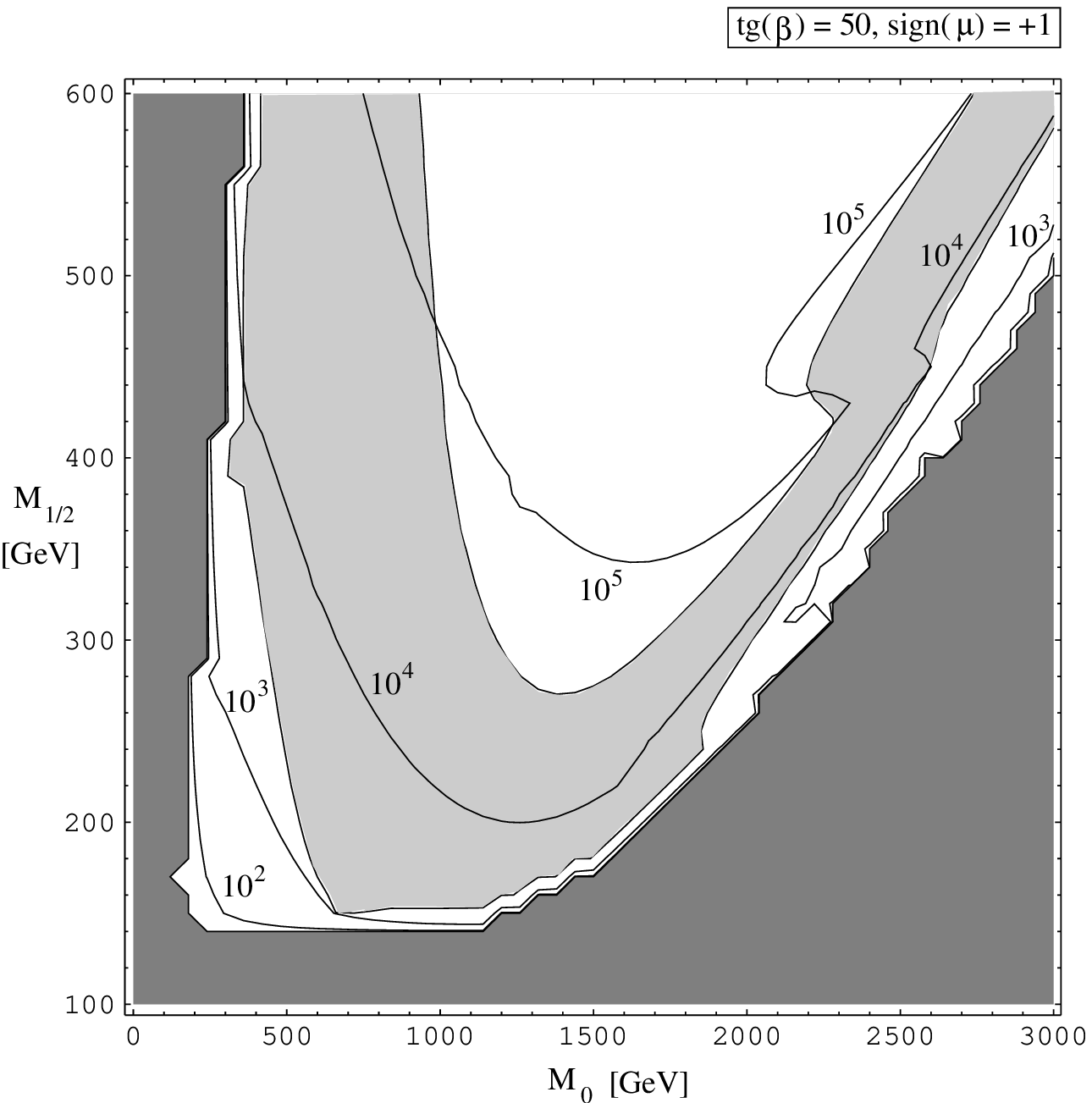}
\caption{Contour plots in the mSUGRA $(m_0,m_{1/2})$ plane, for the value
of the normalization factor $N_\chi$, that allows the detection of the
neutralino $\gamma$ ray signal, with GLAST. The light shaded region
corresponds to the cosmologically favoured region where
$0.1\le\Omega_{\chi} h^2\le 1$, while the dark shaded
one corresponds to models that are excluded either by incorrect EWSB, LEP
bounds violations or because the neutralino is not the LSP.}
\label{msugra-plots}
\end{center}
\end{figure}

The results, for two possible choices of the parameters $\tan\beta$, $A_0$ and $sign(\mu)$, are shown in figure~\ref{msugra-plots}.
The contour plots, in the mSUGRA $(m_0,m_{1/2})$ plane, show the normalization factor $N_\chi$ needed for GLAST detection, at $3\sigma$ level, of the neutralino induced $\gamma$ ray signal.

\section{Conclusions}
We have performed an analysis of the possibility of an indirect detection of supersymmetric dark matter with cosmic $\gamma$-rays. The most important indication is that there is indeed room for supersymmetric dark matter in the already available $\gamma$-ray data from EGRET.
Moreover the computation of the expected $\gamma$-ray flux coming from dark matter annihilation with upcoming detector GLAST shows that it will be possible to study in more details a supersymmetric signal.

\section{Acknowledgements}
This work has been performed in collaboration with Alessandro Cesarini, Francesco Fucito and Aldo Morselli in Roma Tor Vergata University and with Piero Ullio at SISSA in Trieste. I would like to thank the GLAST dark matter working group, in particular the SLAC people, for many interesting discussions on the argument.


\begin{thebibliography}{99}
\bibitem{wmap}
  D.N.~Spergel {\it et al}, astro-ph/0302209.
\bibitem{Haber:1984rc}
H.~E.~Haber {\it et al},
Phys.\ Rept.\  {\bf 117}, 75 (1985).
\bibitem{Griest:kj}
K.~Griest {\it et al},
Phys.\ Rept.\  {\bf 333}, 167 (2000).
\bibitem{Martin:1997ns}
S.~P.~Martin,
arXiv:hep-ph/9709356.
\bibitem{Mayer}  
  H.~Mayer-Hasselwander {\it et al},  Astron. Astrophys. {\bf 335}, 161 (1998).
\bibitem{smapj} 
  A.~W.~Strong {\it et al},
  Astrophys.\ J.\  {\bf 537}, 763 (2000)
  [Erratum-ibid.\  {\bf 541}, 1109 (2000)].
\bibitem{Cesarini:2003nr}
A.~Cesarini {\it et al},
arXiv:astro-ph/0305075.
\bibitem {glast} 
  Proposal for the Gamma-ray Large Area Space Telescope, SLAC-R-522  (1998);
GLAST Proposal to NASA A0-99-055-03 (1999).
\bibitem{nfw}
  J.F. Navarro {\it et al},
  Astrophys.\ J.\  {\bf 462}, 563 (1996). 
\bibitem{moore2}
  S.~Ghigna {\it et al}, Astrophys.\ J.\  {\bf 544}, 616 (2000).
\bibitem{msugra}
   L.~J.~Hall {\it et al},
   Phys.\ Rev.\ D {\bf 27}, 2359 (1983).
\bibitem{Martin:1993zk}
S.~P.~Martin {\it et al},
Phys.\ Rev.\ D {\bf 50}, 2282 (1994)
[arXiv:hep-ph/9311340].
\bibitem{newcoann}
   J.~Edsjo {\it et al},
   JCAP {\bf 0304}, 001 (2003)
   [arXiv:hep-ph/0301106].
\end{thebibliography}
\end{document}